# Drug Repurposing Targeting COVID-19 3CL Protease using Molecular Docking and Machine Learning Regression Approach

Imra Aqeel [1], and Abdul Majid [1]

[1] Biomedical Informatics Research Lab, Department of Computer & Information Sciences, Pakistan Institute of Engineering & Applied Sciences, Nilore, Islamabad 45650, Pakistan; imraaqeel@pieas.edu.pk ; abdulmajiid@pieas.edu.pk

**Abstract:**

The COVID-19 pandemic has initiated a global health emergency, with an exigent need for effective cure. Progressively, drug repurposing is emerging a promise solution as it saves the time, cost and labor. However, the number of drug candidates that have been identified as being repurposed for the treatment of COVID-19 are still insufficient, so more effective and thorough drug exploring strategies are required. In this study, we joint the molecular docking with machine learning regression approaches to find some prospective therapeutic candidates for COVID-19 treatment. We screened the 5903 approved drugs for their inhibition by targeting the main protease 3CL of SARS-CoV-2, which is responsible to replicate the virus. Molecular docking is used to calculate the binding affinities of these drugs to the main protease 3CL. We employed several machine learning regression approaches for QSAR modeling to find out some potential drugs with high binding affinities. Our outcomes demonstrated that the Decision Tree Regression (DTR) model with best scores of $R^2$ and RMSE, is the most suitable model to explore the potential drugs. We shortlisted six favorable drugs with their respective Zinc IDs ZINC (3873365, 85432544, 203757351, 85536956, 8214470, and 261494640) within the range of -15 kcal/mol to -13 kcal/mol. These drugs have novel repurposing potential, except for one antiviral ZINC203757351 compound that has already been identified in other studies. We further examined the physiochemical and pharmacokinetic properties of these most potent drugs and their best binding interaction to specific target protease 3CLpro. Our study provides an efficient framework to explore the potential drugs against COVID-19, and establishes the impending combination of molecular docking with machine learning regression approaches to accelerate the identification of potential therapeutic candidates. Our verdicts contribute to the larger goal of finding effective cures for COVID-19, which is an acute global health challenge. The outcomes of our study provide valuable insights into potential therapeutic candidates for COVID-19 treatment.

## 1 Introduction

The COVID-19 outbreak has presented an unprecedented worldwide health emergency, with more than 704.75 million confirmed cases and over 7.01 million mortalities globally as of 06 June 2024 according to https://www.worldometers.info/coronavirus/. At present, there is no certain drug available to treat COVID-19, and the development of effective cures has become a priority for researchers globally (1). COVID-19 is triggered by SARS-CoV-2, a positive-sense single-stranded RNA virus that mainly infects the respiratory tract of humans (2). When the spike protein attaches to the ACE2 receptor on the surface of human cells, the virus enters the cell, and then it utilizes the host's cellular machinery to replicate and spread throughout the body. Fig. 1 depicts the life cycle of a coronavirus.

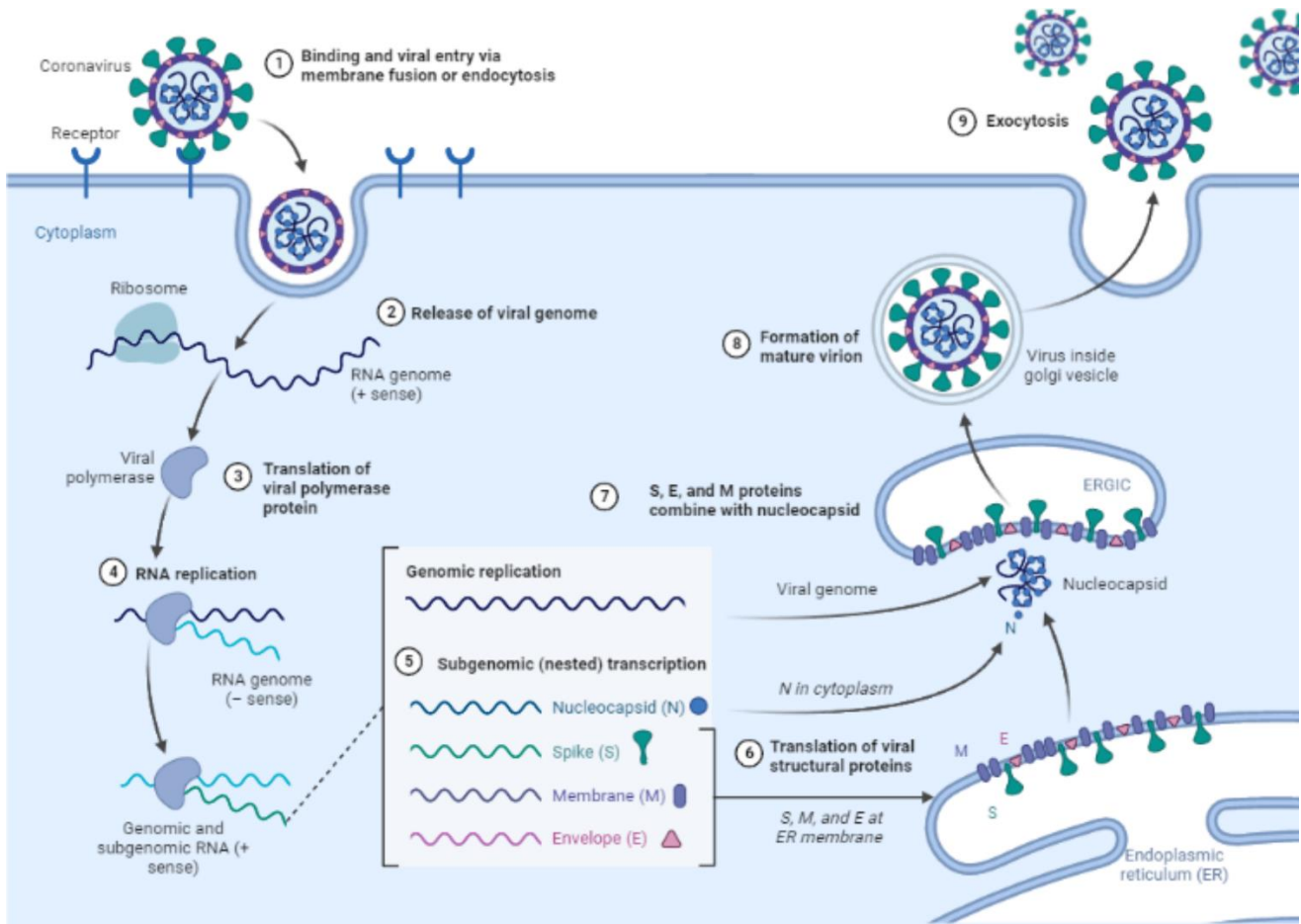

Fig. 1. Life cycle of a coronavirus

To support viral replication, SARS-CoV-2 uses various viral proteins, among which the main protease 3CLpro (also called Main protease Mpro) plays a crucial role in cleaving the viral polyproteins into functional non-structural proteins necessary for viral replication. As a result of its significance in the viral life cycle, 3CLpro has become a potential target for the development of antiviral therapies for COVID-19. The catalytic dyad of His41 and Cys145 in the homodimeric cysteine protease 3CL protease makes it an attractive candidate for the development of protease inhibitors (3). Several studies have reported the successful identification of small molecules and peptides that can effectively inhibit the activity of the 3CL protease in vitro. However, the development of specific and potent inhibitors for the 3CL protease remains a challenge.

In recent years, computational approaches have become increasingly important in drug discovery, especially in the primary stages of drug development. In silico approaches such as molecular docking and machine learning have the potential to speed up drug discovery by screening large numbers of compounds and predicting their potential binding affinities with target proteins. Molecular docking is a commonly used computational method for calculating the binding of small molecules to target proteins. Recently, machine learning algorithms have been progressively employed to improve the accuracy of molecular docking predictions. Regression models have been used specifically to predict binding affinities, which are essential for detecting potential drug candidates. In a recent study (4), a combination of docking, machine learning (ML), and molecular dynamics (MD) calculations was used and seven compounds were recognized as having the potential to inhibit COVID-19 Mpro. In another study (5), researchers employed the techniques of molecular docking and MD simulation and identified the four drug candidates DB07299, DB01871, DB04653 and DB08732 to combat 3CLpro of COVID-19.

The repurposing of drugs for SARS-COV-2 comprises determining novel therapeutic applications for the existing drugs. It has become a good strategy owing to the persuasive need for effective treatments. The approach involves to screen the existing drugs against SARS-CoV-2 targets, with the purpose of identifying compounds that can prevent from viral replication or attenuate the host immune response to the virus. Several studies have used computational methods to identify potential drugs for COVID-19, including 3CL protease inhibitors. For example, in a study (6), through the application of molecular docking and MD simulation methods, the authors identified two hit candidates, namely CMP4 and CMP2, that exhibit a solid interaction with the 3CLpro. In a recent study (7), we employed an hybrid approach of QSAR, ADMET analysis, and molecular docking to detect the potential inhibitors against the 3CL protease of COVID-19. As an outcome, six bioactive molecules were identified, each with a unique ChEMBL ID: 187460, 222769, 225515, 358279, 363535, and 365134. These compounds demonstrate potential as effective inhibitors of the 3CL protease and could serve as promising candidates for further study and development as treatments for COVID-19. Another study (8), utilized virtual screening techniques to repurpose existing drugs for the cure of COVID-19. The authors identified two drugs, lurasidone and talampicillin, as potential candidates. Additionally, they recognized two drug-like molecules using the Zinc database. MD simulation and ADMET analysis were conducted to assess the stability and pharmacokinetic properties of the identified compounds. In a separate study (9), molecular descriptors were computed using random forest, logistic regression, and SVM through a deep learning method. This information was then used in QSAR modeling to estimate the binding affinities of selected drugs with target protease. To develop effective COVID-19 treatments, ML based computational techniques that successfully identify compounds with strong binding affinity have the potential to reduce the cost and time-consuming experiments.

Several recent studies (10), (11), (12), (13), (14), (15) have reported the effectiveness of repurposed drugs against SARS-CoV-2, such as remdesivir, which has got emergency use authorization by the FDA for COVID-19 treatment, as well as some other drugs which have shown promising results in preclinical studies. Conversely, the process of identifying effective repurposed drugs against COVID-19 remains challenging, due to the complex and quickly evolving nature of the disease. Computational approaches, (16), (17) such as molecular docking, have become valuable tools in the identification of potential drug candidates. These approaches allow for the fast screening of large numbers of compounds against specific targets, providing valuable visions into the binding interactions and potential efficacy of the compounds. These studies demonstrate the potential of computational methods in identifying potential drugs for COVID-19, particularly those targeting the 3CL protease. Nevertheless, ML based framework is necessary to accelerate the identification of potential therapeutic candidates.

In this study, we proposed ML based framework for exploring the potential drug candidates against COVID-19. This framework would help to screen the FDA-approved with other world-approved drugs for repurposing as potential COVID-19 cures particularly targeting 3CLpro. Initially, we extracted 5903 drug candidates from the Zinc database. We accomplished molecular docking to calculate the binding affinities of the drugs towards the target protease 3CLpro using a well-known AutoDock-Vina software

(18). To improve the efficiency of drug repurposing approach, we modeled the binding affinities of the drugs towards the target protease 3CLpro using several ML methodologies. Our research highlighted the latent benefits of associating molecular docking with machine learning methodologies. This association provides a powerful approach for the prompt screening and identification of potential drug candidates.

In this proposed work, we selected several ML regression models such as Decision Tree regression (DTR), Extra Trees regression (ETR), Multi-Layer Perceptron regression (MLPR), Gradient Boosting regression (GBR), XGBoost regression (XGBR), and K-Nearest Neighbor regression (KNNR). We used the Zinc database to extract the world-approved including FDA-approved drugs. We performed molecular docking using a well-known AutoDock-Vina software and estimated the binding affinities of the drugs towards the target protease. Their H-bonding and hydrophobic interaction is comprehensively studied. In the next step, 12 diverse types of molecular descriptors are calculated using PaDEL descriptor software (19). Various regression models are built and trained on these diverse types of feature descriptors. The input dataset is divided into two parts in which 80% internal dataset is used for 5-fold cross validation to improve the performance of regression models. The remaining 20% data is used as external dataset for testing these models. The simulated results of regression models are assessed using statistical measures of $R^2$ and RMSE. We observed that our purposed DTR model has improved $R^2$ and RMSE values as compared to other regression models. Further, DTR model has performed better on ten feature descriptors and outperformed on other two feature descriptors of CDK fingerprint and MACCS fingerprint. This highlights that DTR model is most suitable in predicting the binding affinities. Further, we analyzed the physiochemical properties of the shortlisted compounds to understand the behavior of drug molecules in biological systems and determining their efficacy and safety.

The succeeding sections of this paper are arranged as follows: section 2 explains the material and the proposed computational framework employed in this study. The results and the corresponding discussions are presented in section 3. At the end, section 4 outlines the concluding remarks of our investigation.

## 2 Material and Methods

The computational framework proposed in this study is comprised of three modules, as depicted in Fig. 2. Module A encompasses various steps involved in preparing the input dataset. On the other hand, module B illustrates the process of molecular docking, which is used to compute the binding affinities of the drugs with the target protease 3CL. Lastly, module C describes the building of the QSAR model and its performance comparison with different state-of-the-art ML models.

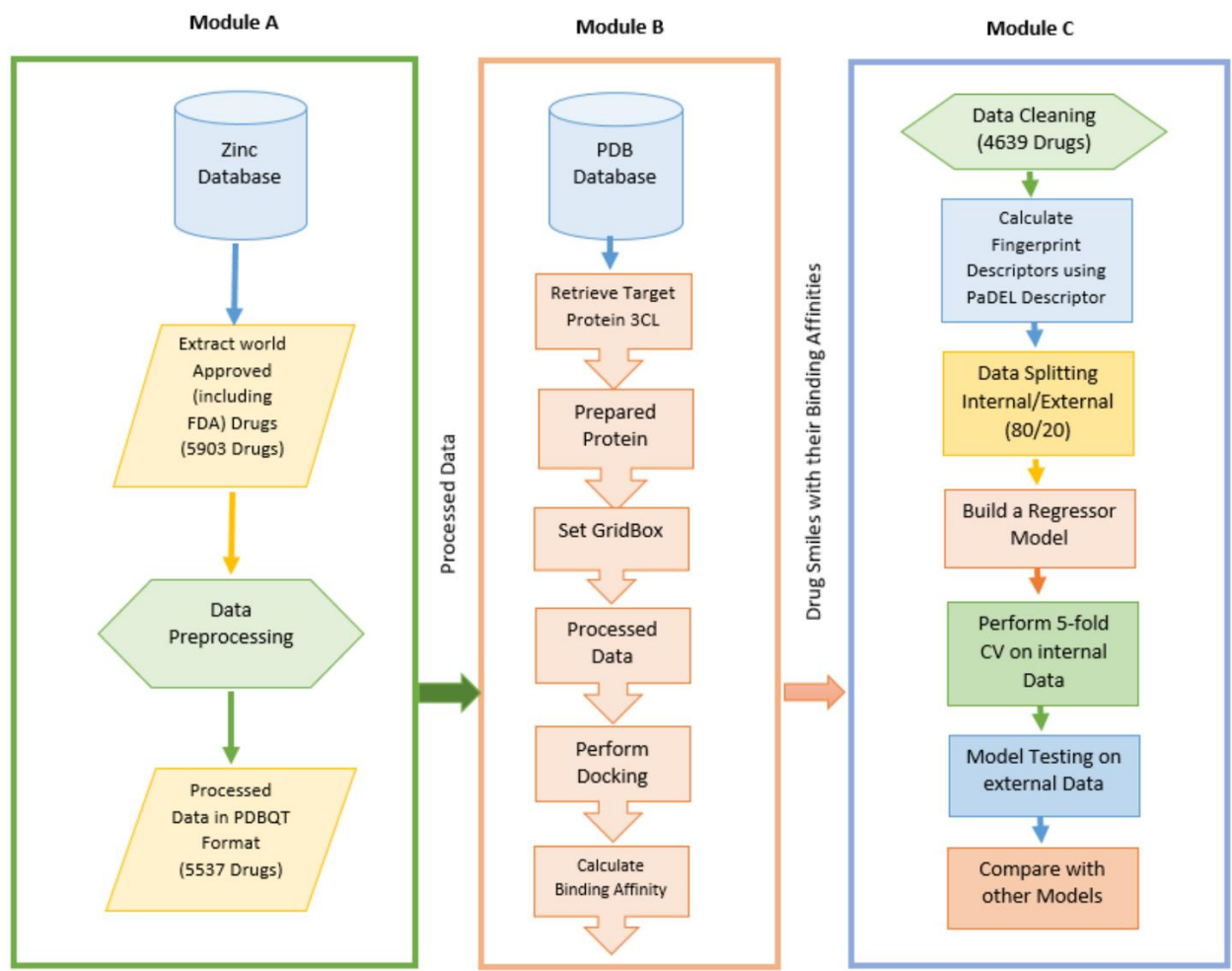

Fig. 2. Three main modules (A to C) in the proposed computational framework

## 2.1. Module A: Dataset Preparation

Module A outlines various stages of data preprocessing, which are as follows:

### 2.1.1. Targeting the Viral Enzyme

The Main protease (Mpro), also known as the 3CLpro is a critical drug target among the proteins of coronaviruses due to its unique enzymatic properties (20). This protease, along with the papain-like protease (PLpro), plays a central role in the transcription and replication process of the viral RNA. This makes it an essential enzyme for the survival and replication of the virus. The high conservation and replication of 3CLpro make it a promising drug target to discover inhibitors that may strongly bind to the target protein and potentially inhibit viral replication.

### 2.1.2. Dataset

The Zinc database is used to extract the world-approved including FDA-approved drugs (21). Initially, a dataset of 5903 drugs was obtained from https://zinc20.docking.org/ on (10/02/2023). The Zinc is publically available database having more than 1.4 billion compounds. Every week data is downloaded from this site in terabytes. More than 90% available compounds are verified.

### 2.1.3. Data Preprocessing

The Zinc dataset consists of 5903 approved drugs that are available in SMILES format. First, these SMILES are converted into SDF format using OpenBabel-2.4.1 software (22). Then, SDF files are transformed into PDBQT format so that these drugs can be used to calculate binding affinities towards the target protease. Those files that could not be converted are dropped. After this preprocessing step, the input dataset is reduced to 5537 drugs.

### *2.2. Module B: Molecular Docking*

The 3CLpro crystal structure (PDB ID: 7JSU) is retrieved from the RCSB Protein Data Bank on 10/02/2023. The ligands were removed from the structure to purify it. The water molecules, and alternative side chains were also eliminated. Further, the polar hydrogen atoms were added and kollman charges were distributed to prepare this macromolecule in a charged form. A GridBox of dimensions 30 x 30 x 30 with a spacing of 01 is fixed to cover the active site of the 7JSU protease, with centers of x, y, and z coordinates are adjusted at −11.046, 12.826, and 67.749, respectively. For molecular docking, Auto-Dock VINA, version 1.2.0 is employed with default parameters (18). Ligands in PDBQT format are prepared using OpenBabel software. The binding affinities are calculated in kcal/mol. The orientation of ligand with target protease interaction having the lowest binding energy is considered the best pose. The crystal structure at the resolution of 1.83 Å of the SARS-CoV-2 3CL protease 7JSU is depicted in Fig. 3.

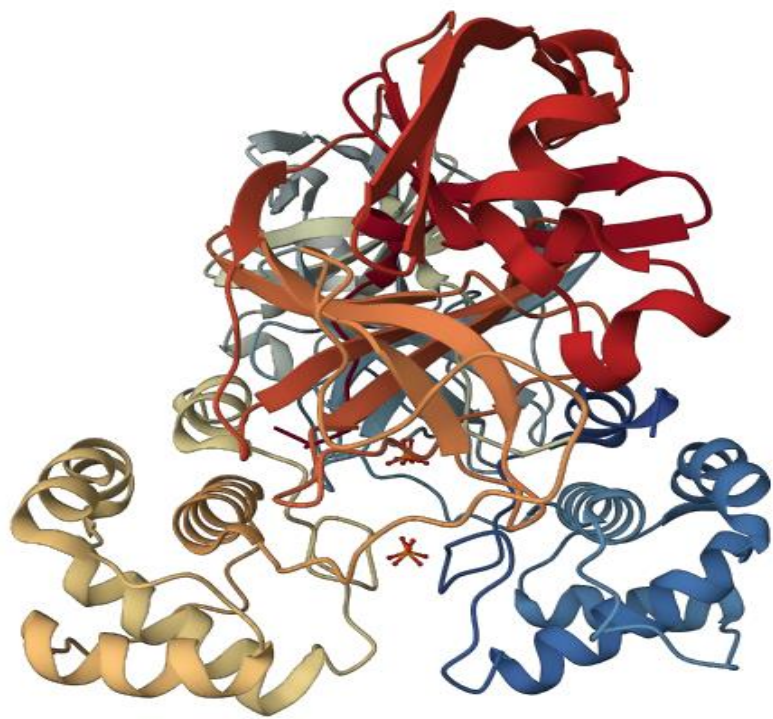

Fig. 3. Crystal structure of SARS 3CL protease 7JSU

*2.3. Module C: QSAR Modeling*

For QSAR modeling, we have selected several ML based regression models such as DTR, ETR, GBR, MLPR, and KNNR. These models predict the quantitative structure-activity relationship (QSAR) between the structural properties of chemical compounds with unknown biological activities. These models have successfully established a correlation between the structural characteristics of known chemical compounds with biological activities. The structural properties denote the physicochemical properties that define the compound's structure, while biological activities represent their pharmacokinetic properties. The molecular descriptors of compounds enable the prediction of how changes in structural characteristics affect biological activity (23).

### 2.3.1. Data Cleaning

To begin the data preparation process for ML models, the dataset undergoes a thorough cleaning procedure to eliminate any instances of data duplication. Additionally, drugs that lack binding affinity values are removed from the dataset to ensure a high quality of data. Following this rigorous cleaning process, the final dataset comprises a total of 4639 drugs that are deemed suitable for subsequent analysis and modeling. The removal of duplicated data and the exclusion of drugs without binding affinity values ensures that the dataset is accurate, reliable, and fit for purpose.

### 2.3.2. Feature Extraction

The molecular constituents of drug molecules were represented by a vector of fingerprint descriptors. Before computing the descriptors, the PaDEL-Descriptor (19) software's built-in function was used to standardize tautomer and eliminate salts. In this study, we investigated the effectiveness of 12 diverse fingerprint descriptors to predict the binding affinities of drug molecules. Table 1 provides a summary of the utilized fingerprints, including their respective size, and description,

Table 1: Summary of Twelve Fingerprint Descriptor Sets.

| Sr. No | Fingerprint Descriptor | Size (Bits) | Description |
|---|---|---|---|
| 1 | CDK | 1024 | Based on its atomic and bond topology, it encodes structural information of a molecule. |
| 2 | MACCS | 166 | The MACCS (Molecular ACCess System) fingerprint is a binary fingerprint representation of a molecule, generated using a predefined set of structural keys. |
| 3 | PubChem | 881 | A binary fingerprint based on PubChem compound database, encrypts molecular substructures and functional groups up to a depth of 4 bonds. |
| 4 | E-state | 79 | E-state fingerprint is a type of molecular descriptor that represents the electronic state of atoms and chemical groups in a molecule. |

| 5 | Extended CDK | 1024 | Based on a predefined set of atom-centered fragments, and having some additional features such as atom types, bonds, and ring sizes that represents the molecular structure. |
|---|---|---|---|
| 6 | 2D Atom Pair | 780 | It encodes the presence of pairs of atoms and their topological distance in a molecule. |
| 7 | 2D Atom Pair Count | 780 | It encodes the frequency of occurrence of atom pairs in a molecule's 2D graph representation. It counts the number of times each atom pair appears in the molecule, and creates a vector of counts for each unique atom pair. The resulting vector represents the 2D atom pairs count fingerprint of the molecule. |
| 8 | Graph Only | 1024 | The Graph Only fingerprint encodes the molecular graph topology, representing the presence or absence of all sub graphs up to a certain size. |
| 9 | Substructure | 307 | It encodes the presence or absence of a predefined set of chemical substructures in a molecule. |
| 10 | Substructure count | 307 | It counts the occurrence of predefined substructures within a molecule to generate a binary vector. |
| 11 | Klekota Roth | 4860 | It encodes the presence and absence of chemical substructures in a molecule. |
| 12 | Klekota Roth count | 4860 | It counts the occurrences of pairs of specific chemical substructures in a molecule. |

**2.3.3. Decision Tree Regression (DTR) Model**

The aim of this research is to develop regression models capable of accurately predicting the continuous response variable, specifically binding affinity, by utilizing a range of predictor variables, such as fingerprint descriptors. To achieve this objective, multiple ML algorithms are developed for QSAR modeling. Among these models, the DTR approach is selected due to its superior prediction performance. In machine learning, a DTR (24), (25) is a predictive model that uses a decision tree to make predictions. The decision tree is a type of graphical model, consisting of nodes, branches, and leaves, that resembles a flowchart. Each internal node in the decision tree represents a test on an attribute, and the outcome of the test is represented by the branch, and each leaf node denotes a prediction or class label. In a DTR, the leaf node shows a continuous value, such as the average or the median of the target values in the training samples that belong to the same leaf node. In DTR, feature space is recursively partitioned into subsets, based on the values of the features in such a manner that maximizes the reduction of the variance of the target variable. This process continues till the stopping criterion is met. It has several advantages (26), such as its interpretability, its ability to handle non-linear relationships between the features and the target variable, and its resistance to over fitting.

The total number of drug molecules in the input dataset, as described in section 2.3.1, is 4639. This dataset is divided into internal and external datasets with 80 to 20 ratio. The internal dataset is used to train and robust the model performance by employing 5-fold cross validation. For this purpose, 12 diverse type of molecular descriptors, describes in section 2.3.2, are used as the feature sets. The external dataset is used to test the performance of the model.

To assess the usefulness of the developed regression models, two statistical variables, namely coefficient of determination ($R^2$) and root mean square error (RMSE), are utilized. The $R^2$ value is a measure of the fraction of variance in the dependent variable that can be described by the independent variables. A value of 0 states a poor fit, while a value of 1 indicates a perfect fit. On the other hand, RMSE provides a measure of the relative error of the predictive model. To compare the performance of different regression models, a comparative analysis is conducted. For Comparative analysis, we utilized two types of fingerprints as the feature sets. One is CDK Fingerprint and the other one is MACCS fingerprint.

## 3   Results and Discussion

In this work, we developed several ML based QSAR models for drug repurposing against COVID-19 and predict their binding affinities for approved drugs towards the target protease 3CLpro. First, we will assess the performance of our proposed model DTR in predicting binding affinities using twelve distinct feature sets. Then compare its performance with several QSAR models using important statistical measures of $R^2$ and RMSE. Then we will explain the results of molecular docking conducted on the world-approved drugs and their interactions with the target protease 3CLpro. Finally, we will conduct the physiochemical analysis of shortlisted drug compounds with respect to the efficacy of the drugs towards the target protease 3CLpro.

### 3.1. Evaluation of QSAR Model

The current research proposes a methodology to construct a QSAR model based on the Decision Tree Regression (DTR) algorithm. The model is developed using a dataset of 4639 drug molecules, which is explained in detail in section 2.3.1. To evaluate the performance of the proposed model, 12 distinct types of feature sets are used, as discussed in section 2.3.2. To construct the data matrices, fingerprint features are placed in the X matrix, while Y matrix consists of their corresponding binding affinities. The dataset is divided into an internal dataset (80%) and an external dataset (20%), where the internal dataset is used to train and robust the model performance by employing 5-fold cross validation. The external dataset is used to assess the model's performance.

The two well-known statistical measures; $R^2$ and RMSE are used to assess the performance of our proposed QSAR model. $R^2$ is used to evaluate the model's fitness and to quantify how much variation in the dependent variable (binding affinity) is explained by the independent variables (features). It ranges between 0 and 1, with higher values indicating better model performance. Conversely, RMSE measures the relative error between the predicted and actual values of binding affinity. To demonstrate the effectiveness of the DTR model, Fig. 4 displays the actual and predicted binding affinity for 12 distinct feature sets.

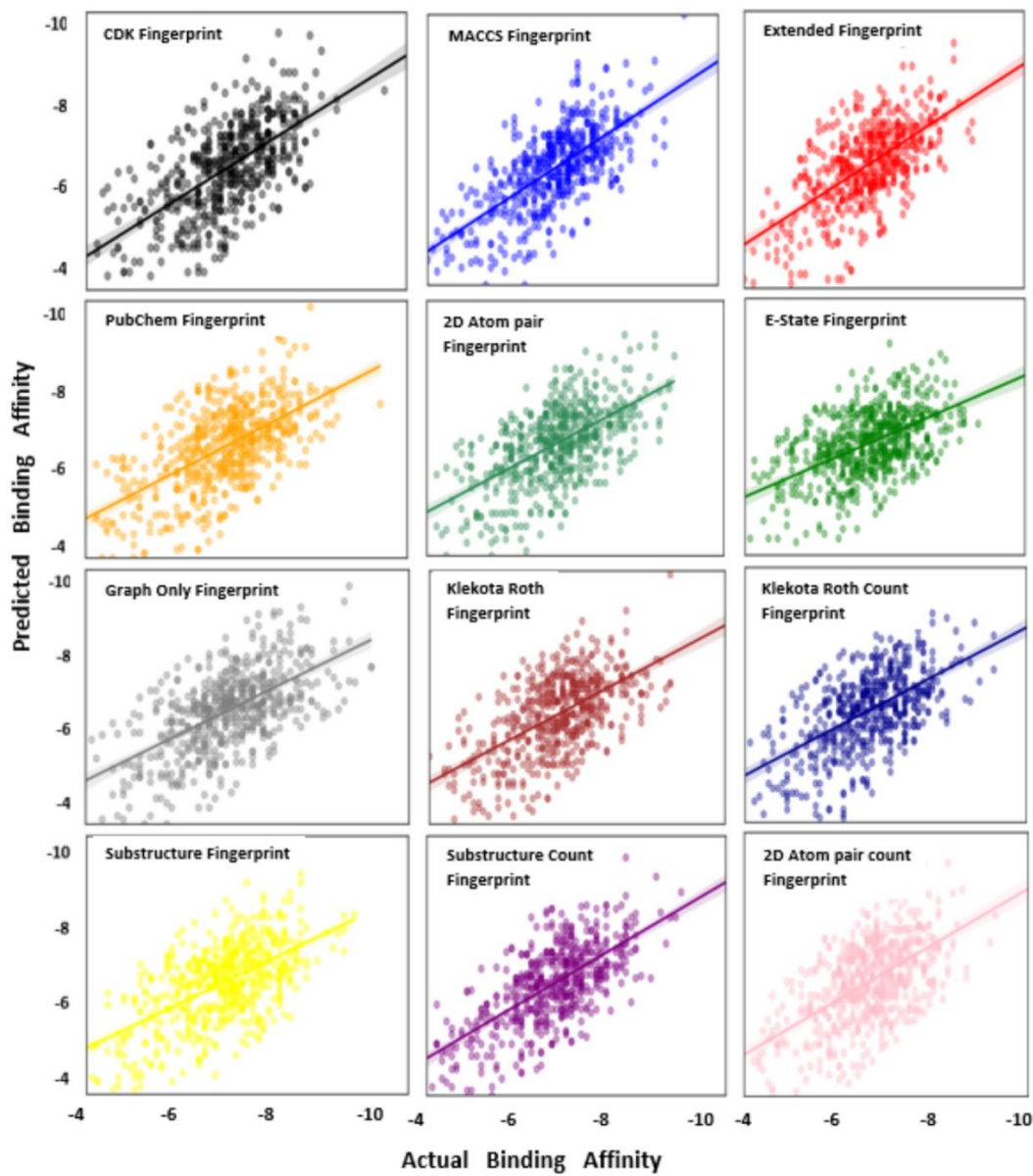

Fig. 4. Scatter plot of 12 feature descriptors for Decision Tree Regression (DTR) model

In this investigation, the effectiveness of the DTR model is analyzed using 12 various feature descriptor sets. The evaluation outcomes, including $R^2$, Mean Squared Error (MSE), and RMSE values, are illustrated in Table 2.

Table 2: Performance of DTR Model for Distinct Fingerprints.

| Sr. No | Descriptors/ Fingerprints | $R^2$ | MSE | RMSE |
|---|---|---|---|---|
| 1 | CDK | 0.97 | 2.81 | 1.68 |
| 2 | MACCS | 0.97 | 2.46 | 1.57 |
| 3 | Extended CDK | 0.90 | 3.38 | 1.84 |
| 4 | PubChem | 0.93 | 3.50 | 1.87 |
| 5 | 2D Atom pair | 0.82 | 2.91 | 1.71 |
| 6 | E-state | 0.70 | 2.70 | 1.64 |
| 7 | Graph Only | 0.81 | 2.87 | 1.69 |
| 8 | Klekota Roth | 0.79 | 3.27 | 1.81 |
| 9 | Klekota Roth count | 0.68 | 3.11 | 1.76 |
| 10 | Substructure | 0.82 | 2.85 | 1.69 |
| 11 | Substructure count | 0.72 | 2.81 | 1.68 |
| 12 | 2D Atom pair count | 0.83 | 3.59 | 1.90 |

Table 2 presents the performance evaluation of 12 different feature descriptors in predicting the binding affinity of a set of drug compounds. Three performance metrics are used to evaluate the model's performance, namely $R^2$, MSE, and RMSE. The $R^2$ metric measures the goodness of fit of the model to the data, where higher values indicate a better fit. The MSE metric calculates the average of the squared differences between the predicted and actual values, while the RMSE metric is the square root of the MSE, and it estimates the error in the same units as the target variable.

The results show that CDK fingerprint and MACCS fingerprint outperformed than other fingerprints with a higher $R^2$ value of 0.97 and a lower RMSE of 1.68 and 1.57, respectively. Achieving an $R^2$ score of 0.93 and an RMSE of 1.87, the PubChem fingerprint displayed strong performance. Other fingerprint descriptors, such as the Extended CDK fingerprint, the 2D Atom pair count fingerprint, the E-state fingerprint, the Graph only fingerprint, the Klekota Roth count fingerprint, the Substructure count fingerprint, and the 2D Atom pair fingerprint, displayed varying degrees of prediction accuracy.

With regard to numerous substructures, functional groups, and molecular features, the CDK fingerprint and MACCS fingerprints provide a thorough description of molecular structure and attributes. They are well suited for exploring a variety of chemical spaces and finding possible candidates for drug repurposing because of their capacity to capture a variety of chemical properties. Taking into account these characteristics, we have chosen the CDK fingerprint and MACCS fingerprints for additional comparison with different ML models while utilizing the proposed DTR model. The data in the table is useful for determining the feature descriptors that will best predict the binding affinity of a group of drug molecules.

### 3.2. Comparative Analysis

A comparison is carried out to assess the predictive performance and accuracy of our proposed QSAR model with other regression models using CDK fingerprint and MACCS fingerprint feature descriptors. The statistical measures $R^2$ and RMSE are used as an evaluators.

Using the CDK fingerprint feature descriptor for the DTR model, Fig. 5 shows a graphical comparison between the actual and predicted binding affinity values. The outcomes demonstrate that the proposed model DTR outperforms the other models with the greatest $R^2$ value of 0.97 for the external dataset. This suggests that the model fits the data well and that the independent variables have a strong correlation with the dependent variable.

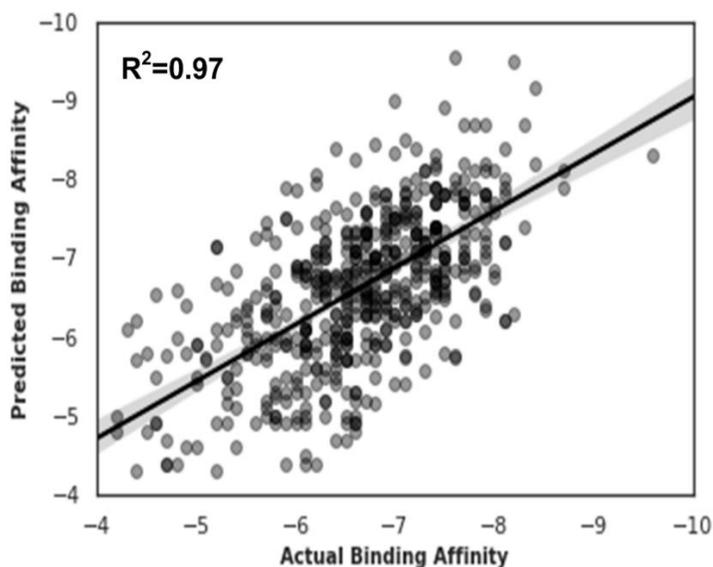
Fig. 5. Regression plot of DTR-QSAR model with CDK fingerprint

The proposed DTR model is also compared with other regression models, using MACCS fingerprint descriptors, to evaluate their performances on the external dataset using $R^2$ and RMSE measures. Fig. 6 presents a graphical description of DTR model using MACCS fingerprint feature descriptors. The results indicated that proposed DTR model outperformed the other regression models, with the highest $R^2$ value of 0.97 for the external dataset. This signifies that our model explained a large quantity of the variance in the dependent variable based on the given set of independent variables, even though the relationship between the variables may not be strictly linear.

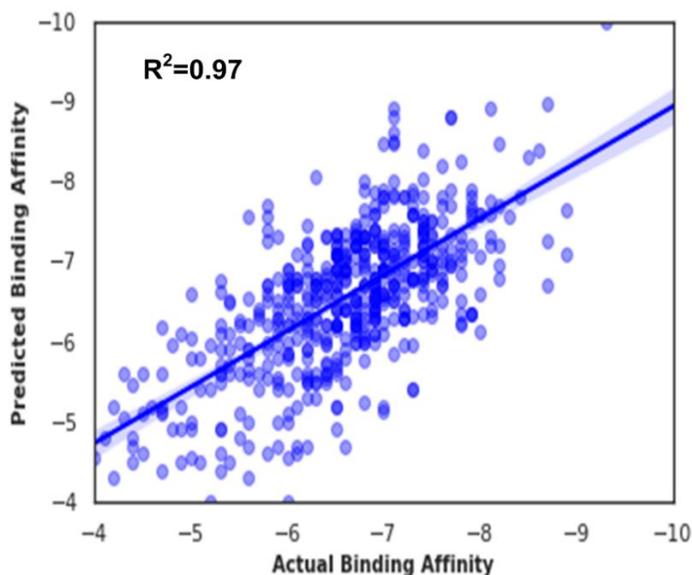
Fig. 6. Regression plots of $R^2$ using MACCS fingerprint features for external dataset

Table 3 highlights the performance comparison of our proposed DTR model with other regression models on two different feature sets, CDK Fingerprint and MACCS Fingerprint in terms of $R^2$, and RMSE values. The DTR achieved $R^2$ value of 0.97 with RMSE value of 1.68 for CDK Fingerprint feature set. For MACCS Fingerprint feature set, DTR model has obtained the $R^2$ value of 0.97, and the RMSE value of 1.57. However, ETR and KNNR models have obtained relatively lower performance for both feature sets. MLPR and XGBR showed poor fit with lower $R^2$ and higher RMSE values for both feature sets. Overall, quantitative results suggest that our proposed DTR model is more suitable choice for predicting the binding affinity for both CDK Fingerprint and MACCS Fingerprint feature sets.

Table 3: Performance Comparison of QSAR Regression Models.

| Regression Model | CDK Fingerprint | | MACCS Fingerprint | |
|---|---|---|---|---|
| | $R^2$ | RMSE | $R^2$ | RMSE |
| DTR | 0.97 | 1.68 | 0.97 | 1.57 |
| ETR | 0.91 | 1.85 | 0.85 | 1.80 |
| KNNR | 0.90 | 1.84 | 0.96 | 1.86 |
| GBR | 0.87 | 1.83 | 0.88 | 1.81 |
| MLPR | 0.70 | 1.74 | 0.60 | 1.64 |
| XGBR | 0.60 | 1.87 | 0.54 | 1.85 |

### *3.3. Molecular Docking*

Our main objective was to determine the efficacy of the selected drug molecules in interacting with the target protease. For this purpose, molecular docking technique used to predict and analyze the interactions between ligand and target protease. It helps in understanding the binding affinity and the orientation of ligands within the protease active site. In our study, we used a ligand-based docking approach to estimate the binding affinities of drug molecules extracted from the Zinc database. The drug molecules were converted into PDBQT format and their binding affinities with the target protein 7JSU were evaluated in kcal/mol units.

Fig. 7 A displays the binding pocket of the target protease 7JSU. On the other hand, Fig. 7 B shows the 3D interaction view of complex of 7JSU with ligand 2297 bound with it. This figure depicts the interacting residues of 7JSU with ligand 2297 atoms along with intermolecular distances. In this interaction, hydrogen bonds are represented by dotted lines shown in green color. A hydrogen bond appears when a hydrogen atom from the protein interacts with an electronegative atom (such as oxygen or nitrogen) from the ligand or vice versa. The distance between the hydrogen donor and the acceptor atom is around 2-3 Å. The distance shows values in the range of 2.36 -2.51 Å for hydrogen bonds. These shorter distances propose that the hydrogen bond interactions between the protein and ligand are relatively strong and stable. The detail numercial description about the hydrophobic interaction and the H-bonding is provided in the next seection. These tables emphases the protein-ligand structural context of hydrophobic contacts, hydrogen bonds, and atomic coordination. These useful information illustrate the structural and energetic aspects of the protein-ligand complex.

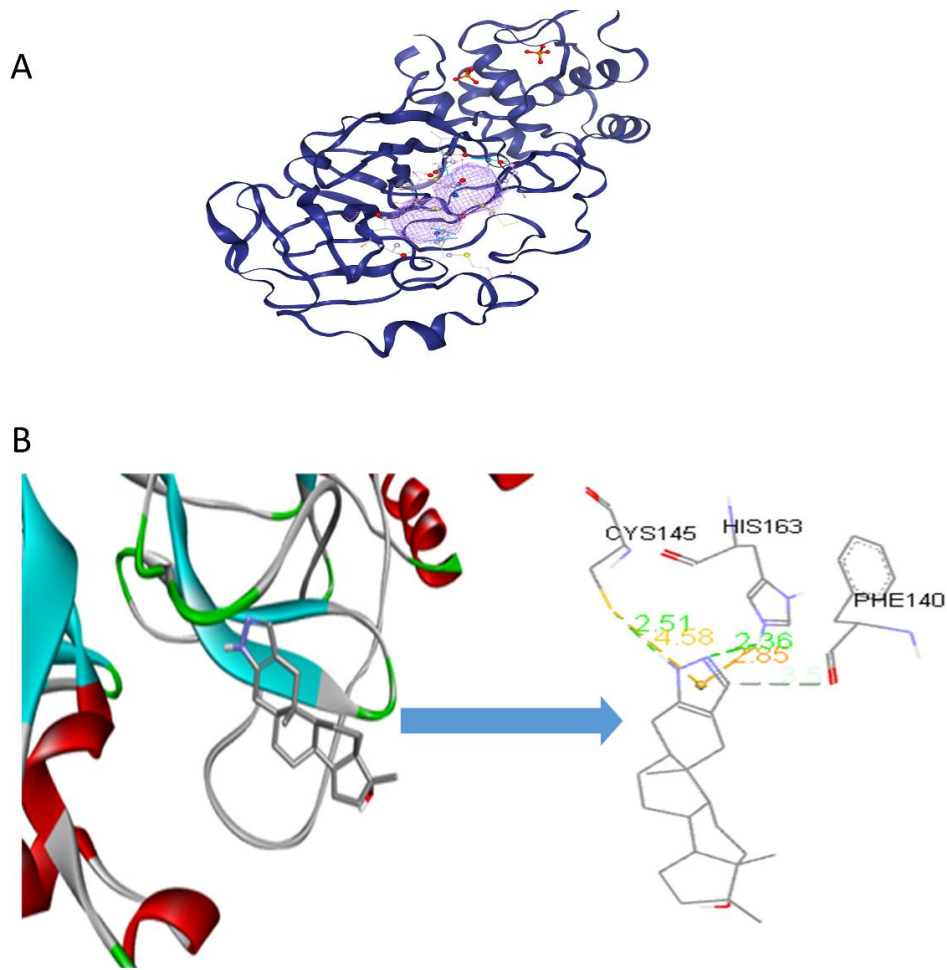

Fig. 7. (A)- The binding pocket of the target protease 7JSU; (B)- 3D interaction view of Protein- ligand ID 2297 Complex.

Fig. 8 demonstrates the 2D view of optimal poses of six top rank drug compounds interacting with the target protease corresponds to the best binding affinity. These drug molecules have the best binding affinities ranging from −15.1 to −13.6 kcal/mol. The optimal pose refers to the specific orientation and conformation of the ligand molecule that achieves the lowest binding affinity with the target protease. The most negative binding energy value represents the best ligand pose towards the target. 2D views show the interaction of protein-ligand in terms of Van Der Waals, conventional hydrogen bond, carbon hydrogen bond, Pi-cation, Pi-sulfur, Pi-Pi T-shaped, unfavorable donor-donor, alkyl, Pi-alkyl. These different types of interactions play important roles in the overall stability of protein-ligand binding.

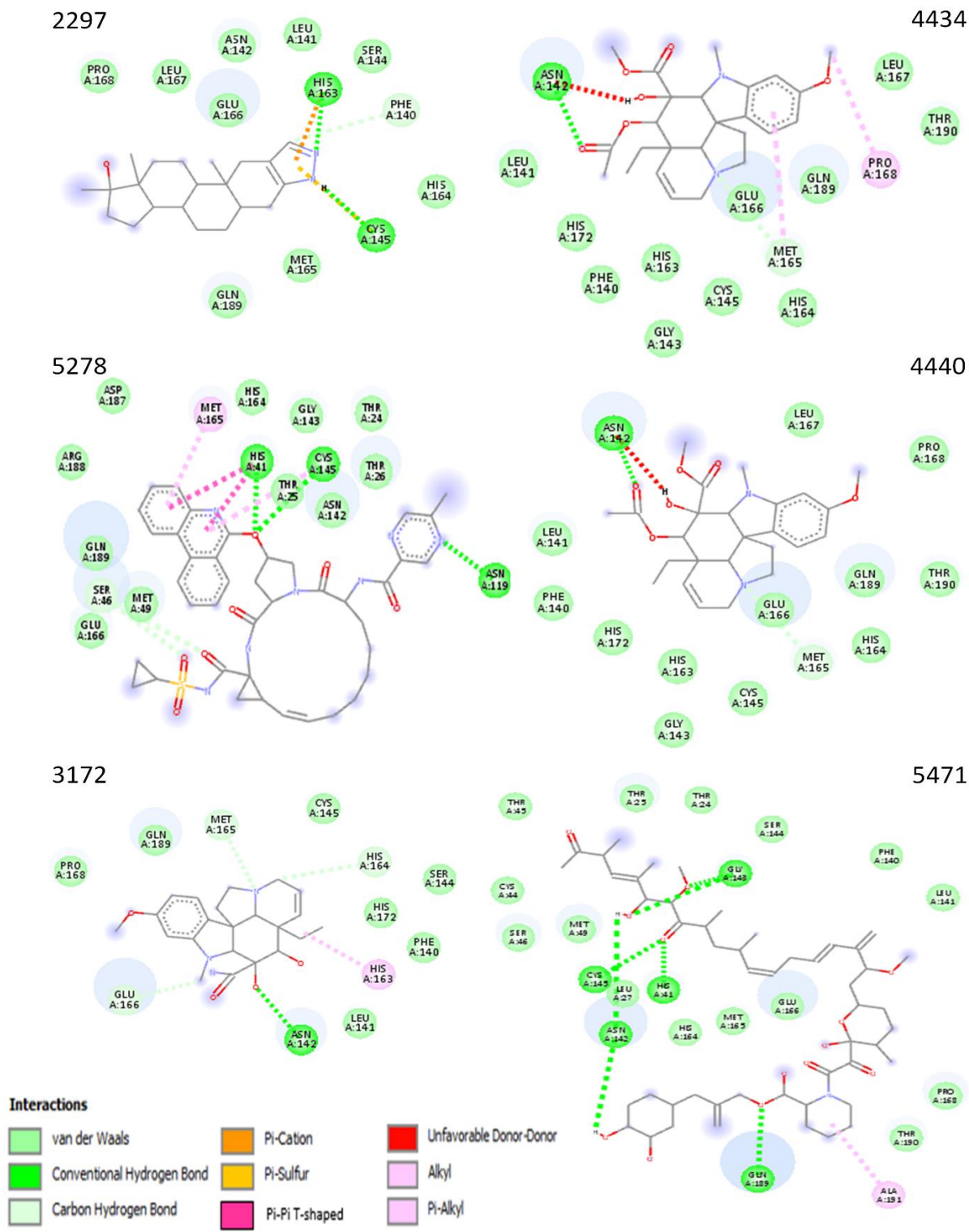

Fig. 8. The optimal poses (2D view) of six top rank drug compounds interacting with target protease 7JSU

On the other hand, docking accuracy is evaluated by measuring the root mean square deviation (RMSD) of the ligand from its original position in the protease complex. A lower RMSD value indicates a superior docking geometry of the ligand molecule. Interestingly, our analysis demonstrates that all six ligands achieved an RMSD value of zero at their optimal poses, implying a high level of accuracy in the docking geometry. Table 4 shows the binding affinities (BA) values of the six top-ranked drug molecules and the corresponding amino acid (AA) residues involved in the hydrophobic interaction and H-bonding. The ligands 2297, 4434, 5278, 4440, 3172, and 5471 have obtained the best BA values -15.1, -14.4, -14.4, -13.9, -13.6, and -13.6 kcal/mol, respectively. This table shows that ligand 2297, with minimum -15.1 kcal/mol, is the most promising drug compound.

The investigation of ligand poses and their corresponding binding energies helps establish a relationship between the molecular structure of the ligands and their binding affinity. It designates the strong interaction and potential efficacy of the ligand molecule as a drug candidate. The identified amino acids involved in hydrophobic interactions and hydrogen bonds offer valuable intuitions into the molecular mechanisms underlying ligand-protein interactions. Further, exploration and analysis of these interactions would contribute to the development of novel therapeutic strategies targeting the specific residues and improving the efficacy of drug candidates.

Table 4: Top Ranked Six Ligands with Target Protein 7JSU.

| Sr. No. | Zinc ID | Ligand ID | BA (kcal/mol) | List of AA Residues | |
|---|---|---|---|---|---|
| | | | | Hydrophobic interaction | H-bonding |
| 1 | ZINC003873365 | 2297 | -15.1 | GLU166 | HIS163 |
| 2 | ZINC085432544 | 4434 | -14.4 | PHE140, ET165, GLU166 | ASN142, GLU166, GLN189 |
| 3 | ZINC203757351 | 5278 | -14.4 | MET165, GLU166, GLN189 | THR26, HIS41, ASN119 |
| 4 | ZINC085536956 | 4440 | -13.9 | PHE140, LEU141, MET165, GLU166 | ASN142, GLU166, GLN189 |
| 5 | ZINC008214470 | 3172 | -13.6 | PHE140, GLU166 | ASN142, GLU166 |
| 6 | ZINC261494640 | 5471 | -13.6 | GLU166, GLN189 | HIS41, ASN142, GLY143, GLN189 |

### 3.3.1. Hydrophobic Interactions

Table 5 provides the hydrophobic Interactions between C-H bonds of six ligands with the chain A of 3CL target protein. Due to non-polar nature, the residues MET, PHE, GLU, GLN, and LEU revealed the hydrophobic interactions. It is worth noting that these hydrophobic residues are concealed within the protein core. The large side chain based hydrophobic residues contribute significantly to the establishment of the protein's hydrophobic core, which plays a key role in maintaining the stability of the ligand-protein structure.

In this table, ligand 2297 exhibits a hydrophobic interaction with residue GLU166A, where a relatively smaller intermolecular distance of 3.22 Å is found between protein atom at position 1578 and ligand atom at position 2889. However, Ligand 3172 give interactions with two amino acid residues of PHE140 and GLU166 with intermolecular distances of 3.62Å and 3.50Å, respectively. Furthermore, this table reveals hydrophobic interactions with other ligands 4434, 4440, 5278, and 5471 with residues MET, PHE, GLU, GLN, and LEU. These Hydrophobic interactions are characterized by varying intermolecular distances ranging from 3.42 Å to 3.95 Å.

Table 5: Hydrophobic Interactions of Top Ranked Six Ligands with the Target Protease.

| Ligand ID | Index | AA Residue | Distance (Å) | Ligand Atom | Protein Atom |
|---|---|---|---|---|---|
| **2297** | 1 | GLU166(A) | 3.22 | 2889 | 1578 |
| **3172** | 1 | PHE140(A) | 3.62 | 2897 | 1336 |
|  | 2 | GLU166(A) | 3.50 | 2897 | 1578 |
| **4434** | 1 | PHE140(A) | 3.94 | 2892 | 1336 |
|  | 2 | MET165(A) | 3.60 | 2890 | 1569 |
|  | 3 | GLU166(A) | 3.06 | 2892 | 1578 |
| **4440** | 1 | PHE140(A) | 3.89 | 2892 | 1336 |
|  | 2 | LEU141(A) | 3.95 | 2869 | 1351 |
|  | 3 | MET165(A) | 3.73 | 2884 | 1569 |
|  | 4 | GLU166(A) | 3.06 | 2892 | 1578 |
| **5278** | 1 | MET165(A) | 3.42 | 2909 | 1569 |
|  | 2 | GLU 166(A) | 3.82 | 2905 | 1578 |
|  | 3 | GLN 189(A) | 3.76 | 2912 | 1788 |
|  | 4 | GLN 189(A) | 3.59 | 2871 | 1787 |
| **5471** | 1 | GLU 166(A) | 3.92 | 2905 | 1578 |
|  | 2 | GLN 189(A) | 3.83 | 2913 | 1788 |

### 3.3.2. H-bonding Interaction

H-bond is a type of intermolecular bond that occurs between H-atom bonded to electronegative N or O atoms. The acceptor N atom within a protein possesses a lone pair of electrons, which interact with H atom from the ligand and vice versa. The donor atom within molecule donates the H-bond. The H-bonds are weaker than covalent bonds. The H-bond interaction analysis help to understand the structural and geometrical stability of protein ligand interaction and to improve the physiochemical and drug biological process. The polar and charged residues in their side chains at different positions such as ASN142, GLU166, THR26, HIS163, and GLY143, etc. The polar residues ASN, GLU, GLN, and THR donate or accept H-bond. The residue HIS has two-NH groups in the side chains, depending on the environment and pH level, can be polar. The residues ARG, LYS and TRP possess N donor atom in their side chains. On the other hand, the residues ASP and GLU are H-bond acceptor (O) atom in side chain.

In the ligand molecules and AA residue combination, the distances H-acceptor and donor-acceptor play a significant role in determining the strength of hydrogen bonds. The smaller distances indicate the stronger H-bonds, as the electrostatic interaction between the partially positive hydrogen and the partially negative acceptor atom is stronger when they are closer together. We compared the values of these distances and analyzed to understand the strength of H-bonding interactions. The optimal (higher) donor angle of each residue is crucial in assessing the strength or weakness of hydrogen bonding. The donor angle represents the spatial orientation of the donor atom involved in the hydrogen bond. The angle influences the alignment and stability of the hydrogen bond, impacting its strength. A more favorable and optimal values of the donor angles results in stronger H-bonds, while deviations from the ideal angle (180) can weaken the interaction. Moreover, the protein donor and acceptor atoms also contribute to the strength of H-bond. These donor and acceptors atoms affect the overall stability and specificity of the hydrogen bonds formed in ligand protein interaction.

The Table 6, highlights the useful values of the H-acceptor distances and donor-acceptor distance, the donor angle, the donor atom, and acceptor atom. These values are focusing on H-A and D-A distances, donor angle, and protein donor/acceptor atoms. The numerical distance values between specific atoms involved in the hydrogen bonding interactions, are given in angstroms (Å). The "Distance H-A" and "Distance D-A" measures represent the "H-Acceptor" distance and "Donor-Acceptor" distance, respectively. These distances provide valuable information about the proximity of the atoms in the hydrogen bond. Their numerical values highlight the strength and geometry of H-bonding interactions in protein-ligand complexes. The specific values of these distances vary depending on the nature of donor and acceptor molecules. The smaller distances indicate the stronger H bonds, which improve the stability of the ligand protein interaction.

For ligand ID 2297, the residue HIS163 forms H-bond with donor angle of 136.68 degrees using the acceptor ligand. The bond is formed between the N+ atoms of the donor protein at 1548 position with N atom of the acceptor at 2887 position. The proximity of the H-acceptor distance (2.36 Å) and donor-acceptor distance (3.18Å) is significant, as the smaller distances indicate the existing of stronger hydrogen bonds for ligand ID 2297. However, for ligand 3172, the ASN142 participates in hydrogen bonding with a donor angle of 120.04 degrees. The donor angle is favorable for the hydrogen bond, enhancing its stability. The H-bond is formed between the protein/donor N atom at position 1359 and the ligand/acceptor oxygen atom at 2893. This introduces more interaction with potential variations in the bonding patterns. For ligand 3172, the optimal smaller values of H-acceptor (3.17 Å) and donor-acceptor (3.78 Å) distances indicate a stronger hydrogen bond.

Similarly, for ligand ID 4434, the residue ASN142 form a donor angle of 122.86 degree with the ligand acceptor. The H-bond is formed between the donor N atom at position 1359 and the ligand/acceptor oxygen atom at 2893 potentially leading to a more diverse hydrogen bonding network. The distances values (2.56 Å and 3.23 Å) of H-acceptor and donor-acceptor signify a hydrogen bond. The donor angle of 122.86 degrees, although different from the ideal 180 angle, still contributes to the stability of the hydrogen bond.

Continuing to ligand ID 4440, the H bond of the residue ASN142 is more significant as compared to residues GLU166, GLN189, and THR26. This residue has a smaller H-acceptor distance of 2.51 Å with larger donor-acceptor distance 4.04Å. This donor angle contributes to the stability and strength of the hydrogen bond, as it aligns the donor and acceptor atoms optimally. The increased donor angle up to 155.73 degree contribute more to the geometrical stability of the H bond. H-bond is formed between the donor atom at 1359 [Nam] and the ligand/acceptor atom at 2898 [O2]. Finally, for ligand ID 5278, the residue THR26 has optimal donor angle of 147.66 degree. The H-bond is formed between the protein donor nitrogen atom at position 224 and the acceptor oxygen atom at position 2889. For this ligand relatively larger distances value 4.07 Å of donor-acceptor help to decrease donor angle up to 147.66 degree, although deviating from the ideal angle (180 degree), still contributes to the stability of the hydrogen bond.

Table 6: H-bonds of Top Ranked Six Ligands with the Target Protease.

| Ligand ID | Index | AA Residue | Distance (Å) H-A  D-A | Donor Angle | Protein donor | Side chain | Donor Atom | Acceptor Atom |
|---|---|---|---|---|---|---|---|---|
| 2297 | 1 | HIS163(A) | 2.36  3.18 | 136.68 | Yes | Yes | 1548 [Npl] | 2887 [N2] |
| 3172 | 1 | ASN142(A) | 3.17  3.78 | 120.04 | Yes | Yes | 1359 [Nam] | 2893 [O2] |
|  | 2 | GLU166(A) | 2.76  3.61 | 140.74 | Yes | No | 1574 [Nam] | 2876 [N3] |
|  | 3 | GLU166(A) | 2.43  3.36 | 159.55 | No | Yes | 2890 [Nam] | 1582 [O-] |
| 4434 | 1 | ASN142(A) | 2.56  3.23 | 122.86 | Yes | Yes | 1359 [Nam] | 2893 [O3] |
|  | 2 | ASN142(A) | 2.48  3.30 | 136.86 | Yes | No | 1353 [Nam] | 2870 [O2] |
|  | 3 | GLU166(A) | 3.01  3.65 | 121.92 | Yes | No | 1574 [Nam] | 2877 [N3] |
|  | 4 | GLN189(A) | 3.45  3.80 | 102.50 | Yes | Yes | 1790 [Nam] | 2899 [O3] |
| 4440 | 1 | ASN142(A) | 3.09  3.37 | 140.66 | Yes | No | 1353 [Nam] | 2870 [O2] |
|  | 2 | ASN142(A) | 2.51  4.04 | 155.73 | Yes | Yes | 1359 [Nam] | 2898 [O2] |
|  | 3 | GLU166(A) | 2.97  3.61 | 121.57 | Yes | No | 1574 [Nam] | 2877 [N3] |
|  | 4 | GLN189(A) | 3.51  3.86 | 102.95 | Yes | Yes | 1790 [Nam] | 2893 [O3] |
| 5278 | 1 | THR26(A) | 3.17  4.07 | 147.66 | Yes | No | 224 [Nam] | 2889 [O2] |
|  | 2 | HIS41((A)) | 2.54  3.28 | 129.13 | Yes | Yes | 380 [Npl] | 2899 [O3] |
|  | 3 | ASN119(A) | 1.99  2.97 | 160.52 | Yes | Yes | 1147 [Nam] | 2920[Nam] |
| 5471 | 1 | HIS41(A) | 1.87  2.74 | 139.95 | Yes | Yes | 380 [Npl] | 2918 [O2] |
|  | 2 | ASN142(A) | 2.93  3.45 | 113.86 | No | Yes | 2931 [O3] | 1360 [O2] |
|  | 3 | ASN142(A) | 2.91  3.73 | 138.46 | Yes | Yes | 1359 [Nam] | 2875 [O3] |
|  | 4 | ASN142(A) | 2.36  3.18 | 139.18 | No | Yes | 2875 [O3] | 1360 [O2] |
|  | 5 | GLY143(A) | 1.92  2.93 | 170.34 | Yes | No | 1364 [Nam] | 2920 [O3] |
|  | 6 | GLN189(A) | 2.43  3.26 | 138.41 | Yes | Yes | 1790 [Nam] | 2893 [O2] |

### 3.4. Physiochemical analysis of drug candidates

On the basis of our analysis, we shortlisted six drug compounds having more efficacy and strong interaction with specific target protease 3CLpro. Table 7 provides a detailed description of these six drug compounds having the lowest binding energies. The description includes the Zinc ID, molecular formula, SMILES, and 2D structure of the drug compounds.

Table 7: Description of Top Ranked Six Drug Compounds.

| Zinc ID | Molecular Formula | SMILES | 2D Structure |
|---|---|---|---|

| ZINC003873365 | C21H32N2O | C[C@]12Cc3cn[nH]c3C[C@@H]1CC[C@H]1[C@@H]2CC[C@]2(C)[C@H]1CC[C@@]2(C)O | 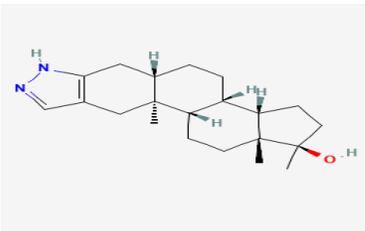 |
|---|---|---|---|
| ZINC085432544 | C46H58N4O9 | CC[C@]1(O)C[C@@H]2CN(CCc3c([nH]c4ccccc34)[C@@](C(=O)OC)(c3cc4c(cc3OC)N(C)[C@H]3[C@@](O)(C(=O)OC)[C@H](OC(C)=O)[C@]5(CC)C=CCN6CC[C@]43[C@@H]65)C2)C1 | 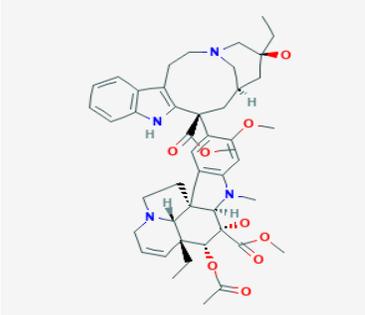 |
| ZINC203757351 | C40H43N7O7S | Cc1cnc(C(=O)N[C@H]2CCCCC/C=C\[C@H]3C[C@@]3(C(=O)NS(=O)(=O)C3CC3)NC(=O)[C@@H]3C[C@@H](Oc4nc5ccccc5c5ccccc45)CN3C2=O)cn1 | 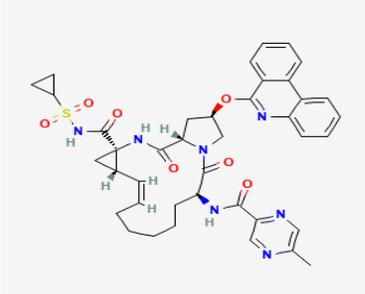 |
| ZINC085536956 | C45H54N4O8 | CCC1=C[C@H]2CN(C1)Cc1c([nH]c3ccccc13)[C@@](C(=O)OC)(c1cc3c(cc1OC)N(C)[C@H]1[C@@](O)(C(=O)OC)[C@H](OC(C)=O)[C@]4(CC)C=CCN5CC[C@]31[C@@H]54)C2 | 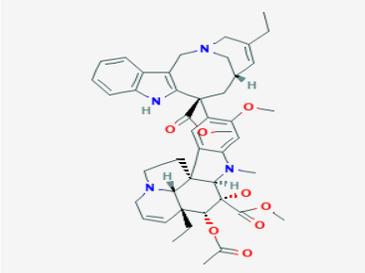 |
| ZINC008214470 | C43H55N5O7 | CC[C@]1(O)C[C@H]2CN(CCc3c([nH]c4ccccc34)[C@@](C(=O)OC)(c3cc4c(cc3OC)N(C)[C@H]3[C@@](O)(C(N)=O)[C@H](O)[C@]5(CC)C=CCN6CC[C@]43[C@@H]65)C2)C1 | 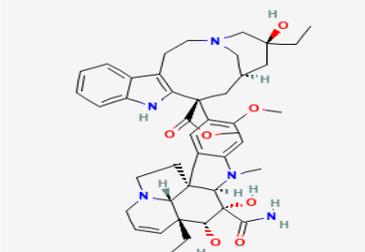 |

| ZINC261494640 | C50H77NO13 | CO[C@H]1C[C@@H]2CC[C@@H](C)[C@@](O)(O2)C(=O)C(=O)N2CCCC[C@H]2C(=O)O[C@H]([C@H](C)C[C@H]2CC[C@H](O)[C@@H](O)C2)CC(=O)[C@@H](C)/C=C(/C)[C@@H](O)[C@H](OC)C(=O)[C@H](C)C[C@H](C)/C=C\C=C/C=C\1C | 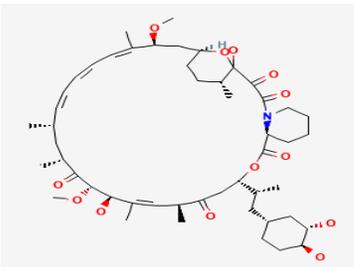 |

Table 8 provides few selected physiochemical properties of six potential drug candidates represented by their Zinc ID, molecular weight, logP, hydrogenbond donors and acceptors, number of rings, heavy atoms, heteroatoms, and fraction sp3. These properties play vital role in the effecacy and bioactivity of the drug compounds. As these properties insight into the potential efficacy and pharmacokinetic profile of the drug, so, these are consider important in drug discovery. To understand the behavior of drug molecules in biological systems and determining their efficacy and safety, the analysis of physicochemical properties is crucial. The binding affinity of drug candidates to certain target proteases is greatly influenced by these properties including molecular weight (Mol.Wt), LogP, rings, hydrogen bond donors (HBD), hydrogen bond acceptors (HBA), heavy atoms, heteroatoms, and the sp3 fraction.

Molecular weight, for instance, is a crucial consideration when assessing the pharmacokinetics and pharmacodynamics of drugs, as well as their capacity to pass across cell membranes and interact with target proteases. The likelihood of binding to target proteases has been found to increase with optimal molecular weight. A similar important component that affects drug binding affinity is logP. The greater lipophilicity, as indicated by a higher logP value, suggests a higher likelihood of interacting with the hydrophobic areas of the protease. A study (27) found that logP was an important factor in calculating the binding affinity of small molecule drugs with target. In the background of drug-protein interactions, HBD and HBA participate in the establishment of specific molecular interactions that are critical for binding and recognition. Another important property, the number of rings in a drug molecule related to binding affinity and selectivity for specific targets. The drug compounds with two or three rings have a higher likelihood of presenting good oral bioavailability and target affinity, while excessively large or complex rings can hinder with binding or increase toxicity (28). The presence of heavy atoms and heteroatoms, particularly nitrogen and oxygen, can pointedly affect drug-target interactions by forming hydrogen bonds or other electrostatic interactions with target residues. However, an imbalance of either too many or too few heteroatoms might impair the solubility of a medicine, membrane permeability, and other vital aspects of bioavailability.

The crucial property the sp3 fraction of a drug compound has been shown to influence its physicochemical properties and target affinity (29). The sp3 fraction signifies the percentage of carbon atoms with three or more single bonds. The higher sp3 fractions are allied with increased water solubility, lower toxicity, and improved pharmacokinetic properties, while extremely low values can lead to poor bioavailability or reduced target selectivity.

Table 8: Physiochemical Properties of Selected Drug Compounds.

| Sr. No | Zinc ID | Mol. Wt (g/mol) | LogP | Rings | HBD, HBA | Heavy Atoms | Hetero Atoms | Fraction sp3 |
|---|---|---|---|---|---|---|---|---|
| 1 | ZINC003873365 | 328.5 | 4.118 | 5 | 2, 2 | 24 | 3 | 0.86 |
| 2 | ZINC085432544 | 810.9 | 3.99 | 9 | 5, 10 | 59 | 13 | 0.59 |
| 3 | ZINC203757351 | 765.8 | 3.637 | 8 | 3, 10 | 55 | 15 | 0.42 |
| 4 | ZINC085536956 | 778.9 | 4.754 | 9 | 4, 9 | 57 | 12 | 0.53 |
| 5 | ZINC008214470 | 753.9 | 2.732 | 9 | 7, 8 | 55 | 12 | 0.58 |
| 6 | ZINC261494640 | 900.1 | 5.527 | 4 | 4, 10 | 64 | 14 | 0.74 |

In the table, the molecular weight of these drugs ranges from 328.5 g/mol to 900.1 g/mol, indicating that the drugs vary broadly in size. We observed that the drugs with higher molecular weight tended to have a higher number of heavy atoms and heteroatoms. This proposes that larger molecules may be more effective in binding to specific targets. However, we also noted that some of the drugs with lower molecular weight have a higher fraction sp3, which may specify a greater degree of three-dimensional complexity and potentially better binding interactions. Further, the logP values, ranges from 2.732 to 5.527, with most of the drugs having a logP value between 3 and 5. Lipophilicity is key in drug development as it can effect drug absorption, distribution, metabolism, and excretion (ADME) properties. Furthermore, most of the values for HBD are ≥ 4 and HBA are ≥ 8 in this table. Higher HBD and HBA in a drug molecule rises its ability to form multiple hydrogen bonds with the protein's binding site. This can improve the interactions and contribute to stronger binding affinity.

This table shows that the number of rings in the molecules varies from 4 to 9 and the heavy atom count ranges from 24 to 64. The stability, potency, and selectivity of a molecule can be impacted by the number of rings and heavy atoms present. Higher complexity and potential for interaction with the target receptor are generally found in molecules with multiple rings and heavy atoms.

According to this table, the range in heteroatom count from 3 to 15, with most of them having between 12 and 15 heteroatoms. These numbers are significant because they can shed light on possible binding interactions between the drug molecules and its specific target. Another key property, the fraction sp3 which denotes the proportion of carbon atoms in the drug that are sp3 hybridized, ranges from 0.42 to 0.90. The most of the drugs having a fraction sp3 value greater than 0.50. A higher fraction sp3 value specifies that a drug has a higher degree of 3D character that can be useful for binding to certain targets.

Overall, these physicochemical properties of selected six drugs play a crucial role in drug discovery and development, and their understanding. These properties are vital for designing drugs with improved efficacy and pharmacokinetic properties. These drug compounds have the potential for repurposing for the treatment of various diseases. These physicochemical properties would be useful in further vitro and in vivo studies that are necessary to determine the efficacy, safety, and dosage of these drugs for the treatment of specific diseases. This useful analysis of these drugs offers a initial point for drug repurposing research. This highlights the worth of considering the physicochemical properties of the drugs for repurposing purposes. Table 9 provides the brief description of our suggested drug candidates to repurpose against COVID-19 3CLpro with their generic name and original medication.

Table 9: Proposed drugs for repurposing in the study

| Sr. No. | Zinc ID | Generic Name | Orignal Purpose/ Treatment | New Indication |
|---|---|---|---|---|
| 1 | ZINC003873365 | Stanozolol | Hereditary angioedema (HAE), anemia, and certain forms of breast cancer. | COVID-19 3CL |
| 2 | ZINC085432544 | Vinblastine | Hodgkin's and non-Hodgkins lymphoma, breast cancer, neuroblastoma, testicular cancer, histiocytosis, mycosis fungoides and Kaposi's sarcoma. | COVID-19 3CL |
| 3 | ZINC203757351 | Paritaprevir | Antiviral agent for Hepatitis C Virus (HCV) infections. | COVID-19 3CL |
| 4 | ZINC085536956 | Vinorelbine tartrate | Advanced or metastatic non-small cell lung cancer. | COVID-19 3CL |
| 5 | ZINC008214470 | Vindesine | Acute leukaemia, malignant lymphoma, Hodgkin's disease, acute erythraemia and acute panmyelosis | COVID-19 3CL |
| 6 | ZINC261494640 | 41-O-demethyl rapamycin | Tumor-based cancers, prevent organ rejection in kidney transplant patients, and coat stents implanted in heart disease patients | COVID-19 3CL |

## 4 Conclusion

In conclusion, our study successfully identified the approved drugs to repurpose for COVID-19 treatment by targeting the SARS-CoV-2 main protease 3CL. Using a computational framework combining molecular docking and machine learning, we identified six drugs with high binding affinities from -15 kcal/mol to -13 kcal/mol and favorable binding energies, indicating their potential as COVID-19 therapeutics. Our proposed Decision Tree Regression (DTR) model outperformed other regression models in predicting binding affinities. Our model has the values of $R^2$=0.97 and RMSE= 1.68 with CDK fingerprints and the values of $R^2$= 0.97 and RMSE=1.57 with MACCS fingerprints. Further, we analyzed the physiochemical properties of the shortlisted compounds to understand the behavior of drug molecules in biological systems and determining their efficacy and safety. Our analysis also demonstrated that the selected compounds effectively inhibit the viral enzyme 3CLpro. These findings highlight the potential of in-silico approaches for drug repurposing and provide insights into the development of potential therapeutic candidates for COVID-19.

Our study puts the foundation for future research endeavors aimed at confirming the effectiveness and safety of the identified compounds as potential therapeutics for COVID-19. Further investigations involving in vitro and in vivo experimentation are warranted to validate their efficacy. Additional studies can also explore the potential synergistic effects of combining these compounds with other existing or novel treatments. Furthermore, our approach can be extended to other viral diseases and even other diseases beyond virology, where drug repurposing can be a viable strategy. Lastly, our study highlights the importance of leveraging computational methods to accelerate the drug discovery process, especially in the face of emerging pandemics and other urgent health crises.

**Funding**: This research did not receive any specific grant from funding agencies in the public, commercial, or not-for-profit sectors.

**Acknowledgment:** The authors would like to acknowledge the support of CompBioAsia 2022, Bangkok Thailand workshop conducted by TEIN* Cooperation Center and National Science Foundation, USA.

**Conflicts of Interest:** There are no conflicts of interest to declare by the authors.